\documentclass[twocolumn,superscriptaddress,longbibliography]{revtex4-1}
\usepackage[usenames,dvipsnames]{xcolor} 
\usepackage[colorlinks=true,urlcolor=blue,citecolor=blue]{hyperref}
\pdfoutput=1
\usepackage{amssymb}
\usepackage{graphicx}
\usepackage[caption=false]{subfig} 

\newcommand{\bra}[1]{\langle {#1} \vert}
\newcommand{\ket}[1]{\vert {#1} \rangle}

\newcommand{\dd}{\textrm{d}}	

\newcommand{\beq}{\begin{equation}}
\newcommand{\eeq}{\end{equation}}

\definecolor{darkgreen}{rgb}{0.0, 0.8, 0.2}

\begin{document}

\title{Quantum Reinforcement Learning: the Maze problem}
\author{Nicola Dalla Pozza}
\affiliation{Scuola Normale Superiore, Piazza dei Cavalieri 7, I-56126 Pisa, Italy}
\affiliation{Department of Physics and Astronomy, University of Florence,\\via Sansone 1, I-50019, Sesto Fiorentino, Italy}
\affiliation{LENS - European Laboratory for Non-Linear Spectroscopy, via Carrara 1, I-50019, Sesto Fiorentino, Italy}
\author{Lorenzo Buffoni}
\affiliation{Department of Physics and Astronomy, University of Florence,\\via Sansone 1, I-50019, Sesto Fiorentino, Italy}
\affiliation{LENS - European Laboratory for Non-Linear Spectroscopy, via Carrara 1, I-50019, Sesto Fiorentino, Italy}
\author{Stefano Martina}
\affiliation{Department of Physics and Astronomy, University of Florence,\\via Sansone 1, I-50019, Sesto Fiorentino, Italy}
\affiliation{LENS - European Laboratory for Non-Linear Spectroscopy, via Carrara 1, I-50019, Sesto Fiorentino, Italy}
\author{Filippo Caruso}
\affiliation{Department of Physics and Astronomy, University of Florence,\\via Sansone 1, I-50019, Sesto Fiorentino, Italy}
\affiliation{LENS - European Laboratory for Non-Linear Spectroscopy, via Carrara 1, I-50019, Sesto Fiorentino, Italy}
\affiliation{QSTAR and CNR-INO, I-50019 Sesto Fiorentino, Italy}

\begin{abstract}
    Quantum Machine Learning (QML) is a young but rapidly growing field where quantum information meets machine learning. Here, we will introduce a new QML model  generalizing the classical concept of Reinforcement Learning to the quantum domain, i.e. Quantum Reinforcement Learning (QRL).
    In particular we apply this idea to the maze problem, where an agent has to learn the optimal set of actions in order to escape from a maze with the highest success probability. 
    To perform the strategy optimization, we consider an hybrid protocol where QRL is combined with classical deep neural networks. In particular, we find that the agent learns the optimal strategy in both the classical and quantum regimes, and we also investigate its behaviour in a noisy environment. It turns out that the quantum speedup does robustly allow the agent to exploit useful actions also at very short time scales, with key roles played by the quantum coherence and the external noise.
    This new framework has the high potential to be applied to perform different tasks (e.g. high transmission/processing rates and quantum error correction) in the new-generation Noisy Intermediate-Scale Quantum (NISQ) devices whose topology engineering is starting to become a new and crucial control knob for practical applications in real-world problems.
    
    This work is dedicated to the memory of Peter Wittek.
\end{abstract}

\maketitle

\section{Introduction}

The broad field of machine learning \cite{bishop_pattern_2011,cover_elements_1991,hastie2009elements} aims to develop computer algorithms that improve automatically through experience with lots of cross-disciplinary applications from domotic systems to autonomous cars, from face/voice recognition to medical diagnostics. Self-driving systems can learn from data, so as to identify distinctive patterns and make consequently decisions, with minimal human intervention. Its three main paradigms are: \emph{supervised learning}, \emph{unsupervised learning} and \emph{reinforcement learning} (RL). The goal of a supervised learning algorithm is to use an output-labeled dataset $\lbrace x_i,y_i \rbrace_{i=1}^N$, to produce a model that, given a new input vector $x$, can predict its correct label $y$. Unsupervised learning, instead, uses an unlabeled dataset $\lbrace x_i \rbrace_{i=1}^N$ and aims to extract some useful properties (patterns) from the single datapoint or the overall data distribution of the dataset (e.g. clustering). In reinforcement learning~\cite{Sutton2018}, the learning process relies on the interaction between an agent and an environment and defines how the agent performs his actions based on past experiences (episodes). In this process one of the main problems is how to resolve the tradeoff between \emph{exploration} of new actions and \emph{exploitation} of learned experience. RL has been applied in many successful tasks, e.g.\ outperforming humans on Atari games~\cite{mnih2015human} and GO~\cite{Silver2016} and recently it is becoming popular in the contexts of autonomous driving~\cite{kiran2020deep} and neuroscience~\cite{botvinick2020deep}.

In recent years lots of efforts have been directed towards developing new algorithms combing machine learning and quantum information tools, i.e. in a new research field known as quantum machine learning (QML) \cite{schuld2015introduction,wittek2014quantum,adcock2015advances,arunachalam2017survey,Biamonte:2017db}, mostly in the supervised \cite{neven2008training, mott2017solving, lloyd20, martina2021machine} and unsupervised domain \cite{otterbach2017unsupervised,vinci2020path,hu2019quantum}, both to gain an \textit{advantage} over classical machine learning algorithms and to \textit{control} quantum systems more effectively. Some preliminary results on QRL have been reported in Refs. \cite{Zonghai2005,Paparo2014} and more recently for closed (i.e. following unitary evolution) quantum systems in Ref.~\cite{Dunjko2016} where the authors have shown quadratic improvements in learning efficiency by means of a Grover-type search in the space of the rewarding actions. During the preparation of this manuscript, Ref. \cite{Saggio2021} has shown how to get quantum speed-ups in reinforcement learning agents where however the quantumness is only in the type of information that is transferred between the agent and the environment. 
However, the setting of an agent acting on an environment has a natural analogue in the framework of open quantum systems~\cite{Breuer2002,CarusoRMP2014}, where one can embed the entire RL framework into the quantum domain, and this has not been investigated in literature yet. Moreover, one of the authors of this manuscript, inspired by recent observations in biological energy transport phenomena \cite{caruso2009highly}, has shown in Ref. \cite{Caruso2016} that one can obtain a very remarkable improvement in finding a solution of a problem, given in terms of the exit of a complex maze, by playing with quantum effects and noise. This improvement was about five orders of magnitude with respect to the purely classical and quantum regimes for large maze topologies. In the same work there results were also experimentally tested by means of an integrated waveguide array, probed by coherent light.

Motivated by these previous works, here we define the building blocks of RL in the quantum domain but in the framework of open (i.e. noisy) quantum systems, where coherent and noise effects can strongly cooperate together to achieve a given task. Then we apply it to solve the quantum maze problem that, being a very complicated one, can represent a crucial step towards other applications in very different problem-solving contexts.

\section{Reinforcement learning}

In RL the system consists of an agent that operates in an environment and gets information about it, with the ability to perform some actions in order to gain some advantage in the form of a reward. More formally, RL problems are defined by a 5-tuple $ (S,A,P_\cdot(\cdot,\cdot),R_{\cdot}(\cdot,\cdot),\gamma)$, where $S$ is a finite set of states of the agent, $A$ is a finite set of actions (alternatively, $A_s$ is the finite set of actions available from the state $s$), $P_a(s,s') = \Pr(s_{t+1}=s' \mid s_t = s, a_t=a)$ is the probability that action $a$ in state $s$ at time $t$  will lead to the  state $s'$ at time $t + 1$, $R_a(s,s')$ is the immediate reward (or expected immediate reward) received after transitioning from state $s$ to state $s'$, due to action $a$, and $\gamma \in [0,1]$ is the discount factor balancing the relative importance of present and future rewards. In this setting one can introduce different types of problems, based on the information one has at disposal. In \emph{multi-armed bandit models}, the agent has to maximize the cumulative reward obtained by a sequence of independent actions, each of which giving a stochastic immediate reward. In this case, the state of the system describes the uncertainty of the expected immediate reward for each action. In \emph{contextual multi-armed bandits}, the agent faces the same set of actions but in multiple scenarios, such that the most profitable action is scenario-dependent. In a \emph{Markov Decision Process} (MDP) the agent has information on the state and the actions have an effect on the state itself. Finally, in \emph{partially observable MDPs} the state $s$ is partially observable or unknown.

\begin{figure}[t]
    \centering
    \includegraphics[width=\linewidth]{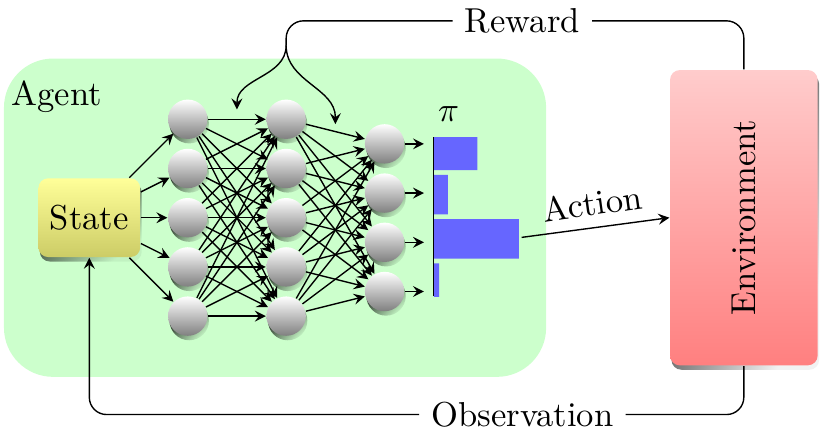}
    \caption{Deep Reinforcement Learning Scheme. A Deep Neural Network learns the policy $\pi$ that the agent uses to perform an action on the environment. A reward and the information about the new state of the system are given back to the agent that improves and learns its policy accordingly.}
    \label{fig:reinforcement}
\end{figure}

The goal of the agent is to learn a policy ($\pi$) that is a rule according to which an action is selected. In its most general formulation, the choice of the action at time $t$ can depend on the whole history of agent-environment interactions up to $t$, and is defined as a random variable over the set of available actions if such choice is stochastic. A policy is called Markovian if the distribution depends only on the state at time $t$, with $\pi_t(a|s)$ denoting the probability to choose the action $a$ from such state $s$, and if a policy does not change over time it is referred as stationary~\cite{Ghavamzadeh2015}. Then, the agent aims to learn the policy that maximizes the expected cumulative reward that is represented by the so-called value function. Given a state $s$, the value function is defined as $V^\pi(s) = \mathbb{E}[\sum_{t=0}^\infty\gamma^tR(Z_t)|Z_0=(s, \pi(.|s))]$, where $Z_t$ is a random variable over state-action pairs. The policy $\pi$ giving the optimal value function $V^*(s) = \sup_\pi V^\pi(s)$ is the RL objective. It is known ~\cite{Sutton2018,Ghavamzadeh2015} that the optimal value function  $V^*(s)$ has to satisfy the Bellman equation, i.e. $ V^\pi(s) = R^\pi(s) + \gamma\int_S P^\pi(s'|s)V^\pi(s')ds'$. In Deep RL, the policy is learned by a Deep Neural Network whose objective function is the Bellman equation itself. The network starts by randomly exploring the space of possible actions and iteratively reinforcing its policy through the Bellman equation given the reward obtained after each action. A pictorial view of this iterative process can be found in Fig. \ref{fig:reinforcement}.

\section{\label{sec:quantum_Maze_and_RL}Quantum Maze}

Here we transfer the RL concepts into the quantum domain where both the environment and the reward process follow the laws of quantum mechanics and are affected by both coherent and incoherent mechanisms. We consider, for simplicity, a quantum walker described by a qubit that is transmitted over a quantum network representing the RL environment. The RL state is the quantum state over the network, represented by the so-called density operator $\rho$. The RL actions are variations of the environment, e.g. its network topology, that will affect the system state through a noisy quantum dynamics. The reward process is obtained from the evolution of the quantum network and hence associated to some probability function to maximize. Following the results in Ref. \cite{Caruso2016} and just to test this framework on a specific model, we consider a \emph{perfect} maze, i.e., a maze where there is a single path connecting the entrance with the exit port. The network dynamics is described in terms of a stochastic quantum walk model \cite{Whitfield2010,Caruso2014}, whose main advantage here is that, within the same model, it allows to consider a purely coherent dynamics (quantum walk), a purely incoherent dynamics (classical random walk), and also the hybrid regime where both coherent and incoherent mechanism interplay or compete each other. Although it is very challenging to make a fair comparison between QRL and RL as applied to the same task and it is out of the scope of this paper, the model we consider here allows us to have the non-trivial chance to analyze the performances of the classical and quantum RL models respectively but in terms of the same resources and degrees of freedom. Very recently we have also exploited this model to propose a new transport-based (neural network-inspired) protocol for quantum state discrimination \cite{DallaPozza2020}.

According to this stochastic quantum walk model, the time  evolution $t$ of the walker state $\rho$ is governed by the following Lindblad equation \cite{lindblad1976generators, Whitfield2010,Caruso2014}: 
\beq
    \frac{d \rho}{d t} = (1-p)\ \mathcal{L}_{QW} (\rho) + p\  \mathcal{L}_{CRW} (\rho) + \mathcal{L}_{exit} (\rho)
    \label{densityEvolution}
\eeq
where $\mathcal{L}_{QW} (\rho) = - i[A,\rho]$ describes the coherent hoping mechanisms, $\mathcal{L}_{CRW} (\rho) = \sum_{i,j} L_{ij} \rho L_{ij}^{\dagger} - \frac{1}{2} \{L_{ij}^\dagger L_{ij}, \rho\}$ with $L_{ij} = (A_{ij}/d_j)\ket{i}\bra{j}$ describes the incoherent hopping ones, while $\mathcal{L}_{exit} (\rho) =  2\ket{n+1}\bra{n}\rho \ket{n} \bra{n+1} - \{\ket{n}\bra{n}, \rho\} $ is associated to the irreversible transfer from the maze (via the node $n$) to the exit (i.e., a sink in the node $n+1$). Here the maze topology is associated to the so-called adjancency matrix of the graph $A$, whose elements $A_{ij}$ are $1$ is there is a link between the node $i$ and $j$, and $0$ otherwise. Besides, $ d_j$ is the number of links attached to the node $j$, while $\ket{i}$ is the element of the basis vectors (in the Hilbert space) corresponding to the node $i$.
The parameter $p$ describes how much incoherent the walker evolution is. In particular, when $p=1$ one recovers the model of a classical random walk, when $p=0$ one faces with a quantum walk, while when $0<p<1$ the walker hops via both incoherent and coherent mechanisms (stochastic quantum walker). Let us point out that the complex matrix $\rho_{ij} \equiv \bra{i} \rho \ket{j}$ contains the node (real) populations along the diagonal, and the coherence terms in the off-diagonal (complex) elements. More in general, in order to have a physical state, the operator $\rho$ has to be positive semi-definite (to have meaningful occupation probabilities) and with trace one (for normalized probabilities). Hence, in this basis only for a classical state $\rho_{ij}$ is a fully diagonal matrix. Then, the escaping probability is measured as $ p_{exit}(t) = 2 \int_0^t \rho_{n n}(t') \ \dd t'$. Ideally, we desire to have $p_{exit}=1$ in the shortest time interval, meaning that with probability $1$ the walker has left the maze.

\begin{figure}[htp]
\includegraphics[width=.95\linewidth]{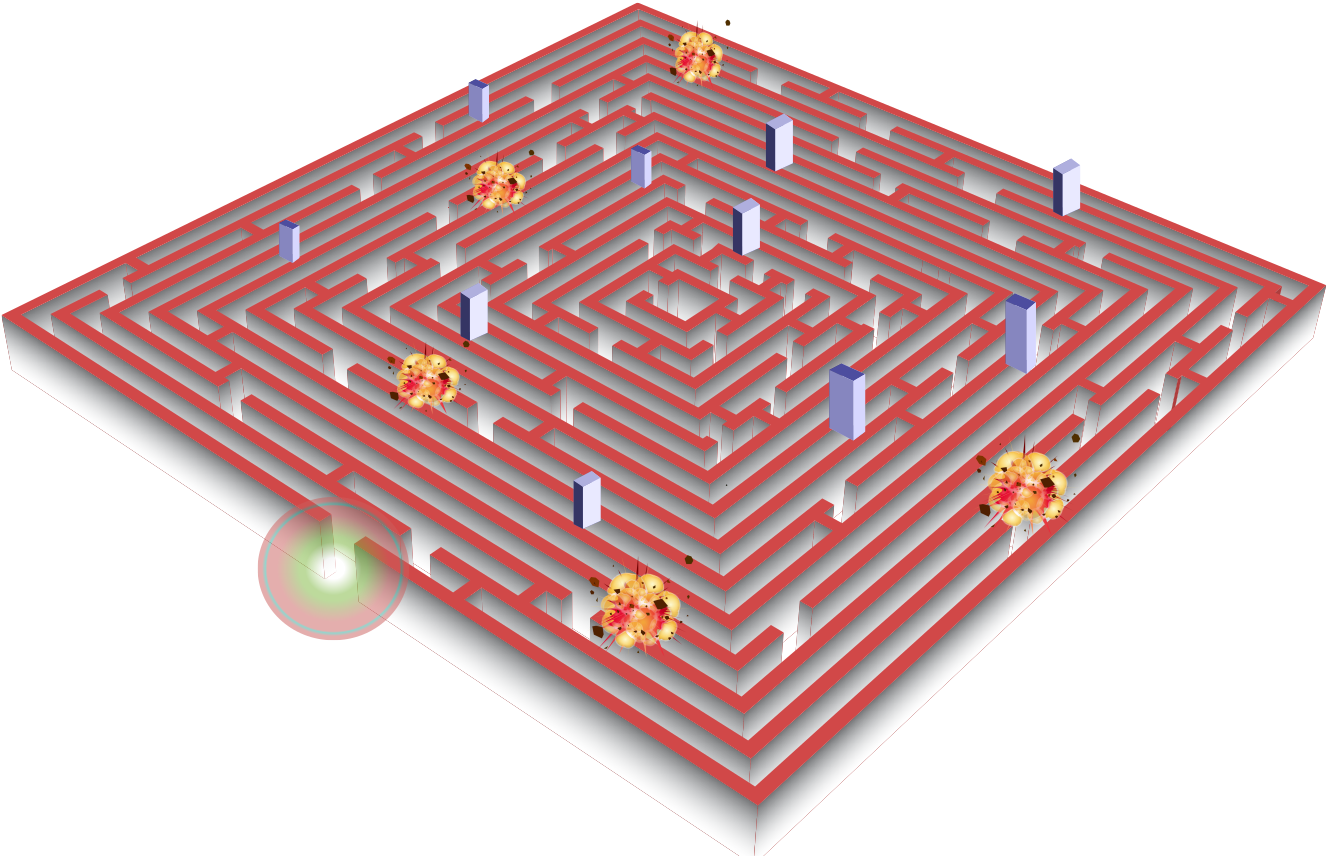}
\caption{Pictorial view of a maze where a classical/quantum walker can enter the maze from a single input port and escape through a single output port. In order to increase the escaping probability withing a certain time, at given time instants the RL agent can modify the environment (maze topology) breaking through existing walls and/or building new walls while the walker moves around to find the exit as quick as possible.}
\label{fig:maze}
\end{figure}

In the RL framework, $\rho(t)$ is the state $s_t$ evolving in time, the environment is the maze, and the objective function is  the probability $p_{exit}$ that the walker has exited from the maze in a given amount of time (to be maximized), or, in an equivalent formulation of the problem, the amount of time required to exit the maze (to be minimized). In this paper we consider the former objective function. The actions are obtained by changing the environment, that is, by varying the maze adjacency matrix. More specifically, we consider three possible actions $a$ performed at given time instants during the walker evolution: (i) building a new wall, i.e. $A_{ij}$ is changed from 1 to 0 (removing a link), (ii) breaking through an existing wall, i.e. $A_{ij}$ is changed from  0 to 1 (adding a new link), (iii) doing nothing (null action) and letting the environment evolve with the current adjacency matrix. The action (i) may allow the walker to waste time in dead-end paths, while the action (ii) may create shortcuts in the maze -- see Fig. \ref{fig:maze}. Notice that the available actions $a$ are indexed with the link to modify, so that the action space is discrete and finite. In the following we set the total number of actions to be performed during the transport dynamics. In principle one could add a penalty (negative term in the reward) in order to let the learning minimize the total number of actions (which might be energy consuming physical processes).
The immediate reward $R_a(s,s')$ is the incremental probability that the walker has left the maze in the time interval $\Delta t$ following the action $a$ changing the state from $\rho(t)$ to $\rho(t+\Delta t)$. This is an MDP setting. The optimal policy $\pi$ gives the optimal actions maximizing the cumulative escaping probability. Besides, one could also optimize the noise parameter $p$ but we have decided to keep it fixed and run the learning for each value of $p$ in the range $[0,1]$.

This approach is slightly different from the scenario pictured 
in the traditional maze problem (classical RL). A classical educational example is provided, for instance, by a mouse (the agent) whose goal is to find the shortest route from some initial cell to a target cheese cell in a maze (the environment). The agent needs to experiment and exploit past experiences in order to achieve its goal, and only after lots of trials and errors it will solve the maze problem. In particular, it has to find the optimal sequence of states in which the accumulated sum of rewards is maximal, for instance considering a negative reward (penalty) for each move on free cells in the maze. This is indeed an MDP setting, where the possible actions are the agent moves (left, right, up, down). In our case we face instead with a probability distribution to find the walker on the maze positions, while in the classical setting the corresponding state would be a diagonal matrix $\rho_{ii}$ where only one element is equal to $1$ and the others are vanishing. Our setup introduces an increased complexity with respect the classical case, in both the definition of the state and in the number of available actions. In addition, a quantum walker can move in parallel along different directions (quantum parallelism), as due to the quantum superposition principle in quantum physics, and interfere constructively or destructively on all maze positions, i.e. the quantum walker behaves as an electromagnetic or mechanical wave travelling through a maze-like structure (wave-particle duality). For these reasons it is more natural to consider topology modifications (i.e. in the hopping rates described by $A_{ij}$) as possible actions. However, let us point out that changing the hopping rate is qualitatively similar to the process of forcing the walker to move more in one or in the other direction, hence mimicking the continuous-version of the discrete moves for the mouse in the classical scenario.

\section{Results}

Within the setting described above, we set a time limit $T$ for the overall evolution of the system and define the time instants $t_k = k \tau$, with $\tau = \Delta t=T/N$ and $k=0, \ldots N-1$, when the RL actions can be performed. The quantum walker evolves according to the Eq. (\ref{densityEvolution}) in the time interval between $t_k$ and $t_{k+1}$.
\begin{figure}[t]
    \centering
    \includegraphics[width=\linewidth]{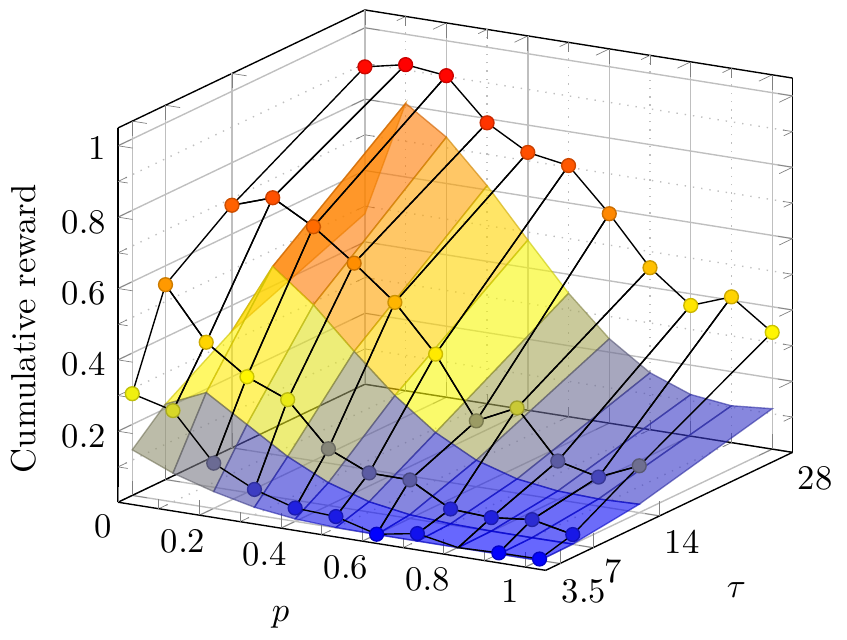}
    \caption{Cumulative reward as a function of $p$ and $\tau$, for a given $6 \times 6$ perfect maze and $N=8$ actions, equally spaced in time by the amount $\tau$. The time unit is given in terms of the inverse of the sink rate set to $1$. The dotted grid above represents the performance of the quantum walker after the training, while the colored solid surface below is the baseline on the same maze with no actions performed by the agent (only free evolution). Repeating the training on over 30 random $6 \times 6$ mazes and averaging their performances for each $(p, \tau)$ we qualitatively obtain the same trend.}
    \label{fig:surface}
\end{figure}
We then implement deep reinforcement learning with $\varepsilon$-greedy algorithm for the policy improvement, and run it with $N=8$ actions and with 1000 training epochs (see Methods for more technical details). At each time instant $t_k$ the agent can choose to modify whatever link in the maze, albeit we would expect its actions to be localized around the places where it has the chance to further increase the escaping probability. The $\varepsilon$-greedy algorithm implies that the agent picks either the action suggested by the policy with probability $1-\varepsilon$ or a random action with probability $\varepsilon$. This method increases the chances of the policy to explore different strategies searching for the best one instead of just reinforcing a sub-optimal solution. The value of $\varepsilon$ is slowly decreased during training so that, at the end, the agent is just applying the policy without much further exploration. This optimization is repeated for different values of $p$ and $T$ in order to investigate their role in the learning process.

\begin{figure}[htbp]
    \centering
    \includegraphics[width=\linewidth]{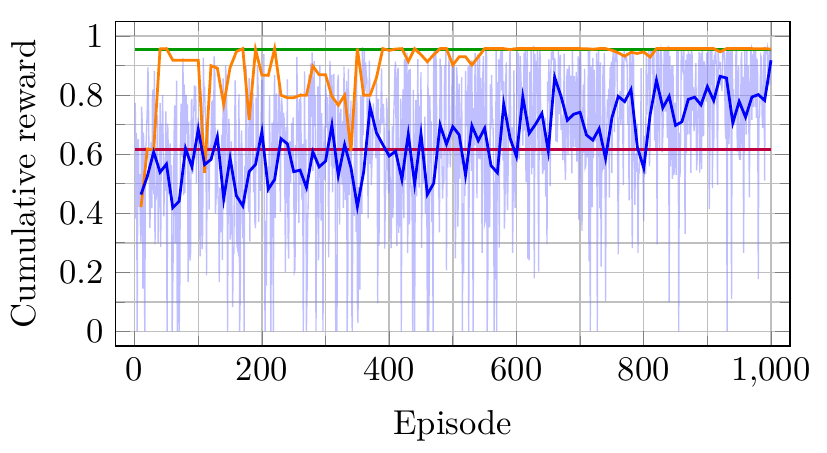}
    \caption{Training curves for an agent performing RL actions for $p=0.4, \tau=28$, and $N=8$ actions on a $6 \times 6$ perfect maze. The curves show the cumulative rewards from single episodes (light blue), ten-episodes window average (dark blue) and for the target network (orange) -- see RL optimization in the Methods. The two horizontal lines are the (constant) cumulative reward in the case of no RL actions (magenta) and for the final trained policy (green).
    \label{fig:training_curves}
    }
\end{figure}

As shown in Fig. \ref{fig:surface}, there is a clear RL improvement for any value of $p$ especially for large $T$ (i.e. also large $\tau$), while for small $T$ it occurs mainly in the quantum regime (i.e. $p$ going to 0) when the walker exploits coherent (and fast) hopping mechanisms. This is due to the fact that the classical random walker (without RL) moves very slowly and remains close to the starting point for small $T$, as reported in Ref. \cite{Caruso2016}. Repeating this experiment for 30 random $6\times 6$ perfect mazes, we find a very similar behaviour -- see also Fig. \ref{fig:mean_surface} in the Methods where interestingly a dip in the cumulative reward enhancement is shown at around $p=0.1$ where the interplay between quantum coherence and noise allows to optimize the escaping probability without acting on the maze \cite{Caruso2014}. There it was very remarkable to observe that a small amount of noise allows the walker to both keep its quantumness (i.e. moving in parallel over the entire maze) and learn the shortest path to the exit from the maze.
Fig. \ref{fig:training_curves} shows an example of cumulative rewards obtained from the training of a network while the agent explores the space of the possible actions. Initially some random actions are performed, and soon the agent finds some positive reinforcement and learns to consistently apply better actions outperforming the case with no actions. 

\begin{figure}[htbp!]
    \centering
    \includegraphics[width=\linewidth]{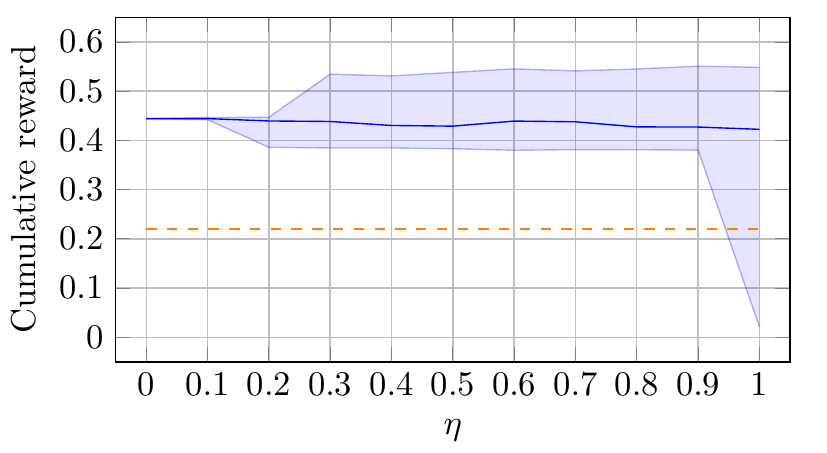}
    \caption{Cumulative reward of an agent trained at $\tau=14$ and $p=0.4$ and deployed in a stochastic environment controlled by the parameter $\eta$. The solid line is the average reward obtained by the agent over $100$ realizations of the noise, the shaded area represents the minimum and the maximum achieved reward and the dashed orange line is the baseline of the walker with no actions performed. While the average performance of the agent remains stable, the variance in the outcomes increases greatly as $\eta$ increases.}
    \label{fig:noise_variance}
\end{figure}

The proposed way of \emph{learning} the best actions also comes with an intrinsic robustness to stochastic noise. Indeed this is a crucial property of RL-based approaches. In our case, we can suppose that we do not have perfect control on the system and there might be perturbations, for example in the timing at which the actions are effectively performed. These kind of perturbations are in general detrimental for hard-coded optimization algorithms, and we want to analyze how our QRL approach performs in this regard. To check this, we first train the agent in an environment with fixed $\tau$ and $p$.
Afterwards, we evaluate the performance of the trained agent in an environment where the time $\tau$ at which the actions are performed becomes stochastic (noisy). This additional noise in the time is controlled by a parameter $0 \leq \eta \leq 1$ while the total time of the actions is kept fixed. In this setting we observe a remarkable robustness of our agent that is capable of great generalization and keep the cumulative reward almost constant despite of the added stochasticity.
Indeed, in Fig.~\ref{fig:noise_variance} we plot the average reward obtained by the agent in this stochastic environment over $100$ different realizations of the noise. We can see that as we increase the parameter $\eta$ our agent, on average, keeps the ability to find the correct actions in order to make the reward consistent even in a stochastic environment, even though it has not been retrained in the noisy setting.
However, while the average reward remains stable, the difference between the minimum and maximum reward increases significantly as $\eta$ increases.

The other tested scenario is the one in which, instead of taking the actions equally spaced in the total evolution time, we concentrate them all at the beginning or at the end of the evolution. This gives our agent a different environment at which it adapts once again implementing different strategies. Indeed, we find that our training method is applicable with success also in this more general scenario thus concluding our remarks on the robustness of the proposed QRL implementation.

A detailed discussion of the robustness analysis and all the aforementioned experiments can be found in the SI.

\section{Discussion}

To summarize, here we have introduced a new QML model bringing the classical concept of Reinforcement Learning into the quantum domain but also in presence of external noise. An agent operating in an environment does experiment and exploit past experiences in order to find an optimal sequence of actions (following the optimal policy) to perform a given task (maximizing a reward function). In particular this was applied to the maze problem where the agent desires to optimize the escaping probability in a given time interval. The dynamics on the maze was described in terms of the stochastic quantum walk model, including exactly also the purely classical and purely quantum regimes. This has allowed to make a fair comparison between transport--based RL and QRL models exploiting the same resources.
We have found that the agent always learns a strategy that allows a quicker escape from the maze, but in the quantum case the walker is faster and can exploit useful actions also at very short times. Instead, in presence of a small amount of noise the transport dynamics is already almost optimal and RL shows a smaller enhancement, hence further supporting the key role of noise in transport dynamics. In other words, some decoherence effectively reproduces a sort of RL optimal strategy in enhancing the transmission capability of the network. Moreover, the presence of more quantumness in our QRL protocol leads to have more robustness in the optimal reward with respect to the exact timing of the actions performed by the agent.

Finally, let us discuss how to possibly implement the RL actions in the maze problem from the physics point of view. In Ref.~\cite{Caruso2014} one of us has shown that one can design a sort of noise mask that leads to a transport behaviour as if one had modified the underlying topology. For instance, dephasing noise can open shortcuts between non-resonant nodes, and Zeno-like effects can suppress the transport over a given link, hence mimicking the two types of RL actions discussed in this paper. As future outlooks, one could test these theoretical predictions via atomic or photonic experimental setups or even on the new-generation NISQ devices whose current technologies today allow to engineer complex topologies and modify them in the same time scale of the quantum dynamics while also exploiting some beneficial effects of the environmental noise that cannot be suppressed.

\section*{Acknowledgments}
This work has much benefited from the initial contribution of Peter Wittek and very stimulating discussions with him before he left us, and is dedicated to his memory. In February 2019, during a workshop on 'Ubiquitous Quantum Physics: the New Quantum Revolution I' at ICTP in Trieste, Peter and F.C. had indeed started a new collaboration focused on the development of the new concept of quantum reinforcement learning that, unfortunately alone, we have carried out later and finally discussed in this paper.

This work was financially supported from Fondazione CR Firenze through the project QUANTUM-AI, the European Union’s Horizon 2020 research and innovation programme under FET-OPEN Grant Agreement No.\,828946 (PATHOS), and from University of Florence through the project Q-CODYCES.

\section*{Authors contribution}
N.D.P. and F.C. led and carried out the theoretical work. N.D.P., L.B. and S.M. performed the numerical simulations and optimizations. F.C. conceived and supervised the whole project. All authors contributed to the discussion, analysis of the results and the writing of the manuscript.

\section*{Methods}

\subsection*{Quantum Maze Simulation}
To simulate the Stochastic Quantum Walk on a maze we have used the popular QuTiP package \cite{johansson2012qutip} for Python. In order to account for the actions performed by the agent at time instants $t_k=k \tau$ modifying the network topology and to evaluate the reward signal, we have wrapped the QuTiP simulator in a Gym environment \cite{brockman2016openai}. Gym is a python package that has been created by OpenAI specifically to tackle and standardize reinforcement learning problems. In this way we can apply any RL algorithms on our quantum maze environment. The initial maze could be randomly generated or loaded from a fixed saved adjacency matrix in order to account for both the  reproducibility of single experiments and the averaging over different configurations.

\subsection*{RL optimization}

\begin{figure}[htbp]
    \centering
    \includegraphics[width=\linewidth]{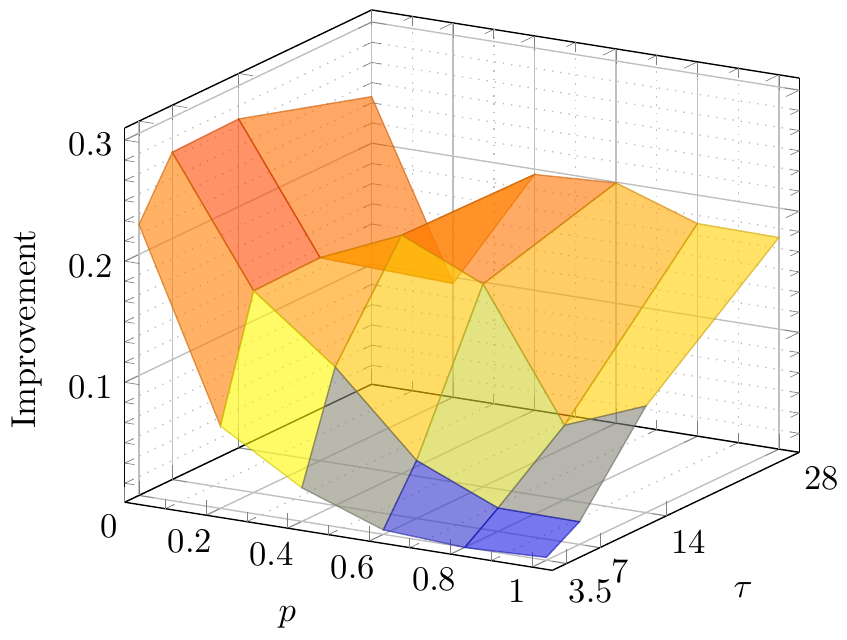}
    \caption{Cumulative reward improvement over the no RL action (free evolution) dynamics as a function of $p$ and $\tau$, averaged over 30 random perfect mazes ($6 \times 6$ size) and $N=8$ (equally spaced in time) actions.}
    \label{fig:mean_surface}
\end{figure}

We have used a feed-forward Neural Network to learn the policy of our agent, following the Deep Q Learning approach \cite{Stooke2019}, realized with the PyTorch package for python \cite{PyTorch}. In this approach, at each iteration of the training loop defining a training epoch, a new training episode is evaluated by numerically solving Eq. (\ref{densityEvolution}) for the time evolution and employing an $\varepsilon$-greedy policy for the action selection. The new training episode is recorded in a fixed-dimension pool of recent episodes called replay memory, from which, after every new addition, a random batch of episodes is selected to train the policy neural network. The $\varepsilon$ parameter is reduced at each epoch, in order to reduce the exploration of new action sequences and increase the exploitation of the good ones proposed by the policy neural network. Periodically, the policy neural network is copied in a target neural network, i.e. a trick used to reduce the instabilities in the training of the policy neural network.  Fig.~\ref{fig:training_curves} shows the reward of the training episodes, their ten--episodes window average, the reward provided by target network, alongside the free evolution (no RL actions) and final reward (constant lines) provided by the trained target network.
Despite the relative simple architecture, we have found the training to be quite sensitive to the choice of learning hyper-parameters, such as the batch size of the training episodes per epoch, the replay memory capacity, the rate of target network update and the decay rate of $\varepsilon$ in the $\varepsilon$-greedy policy. In particular, in Fig. \ref{fig:surface} for each $(p,\tau)$ we run multiple independent hyper-parameter optimizations and training, employing the libraries Hyperopt \cite{Hyperopt} and Tune \cite{liaw2018tune}. Due to the small size of the networks, we were able to launch multiple instances of our training procedure using a single Quadro K6000 GPU. Figure \ref{fig:mean_surface} shows the mean cumulative reward improvement between the no--action strategy (only free evolution) and the trained strategy over 30 random perfect mazes (size $6 \times 6$) with $N=8$ actions.

\section*{Supplementary Information}

\subsection*{Generalization of the Neural Network Training}

\begin{figure}
    \centering
    \includegraphics[width=\linewidth]{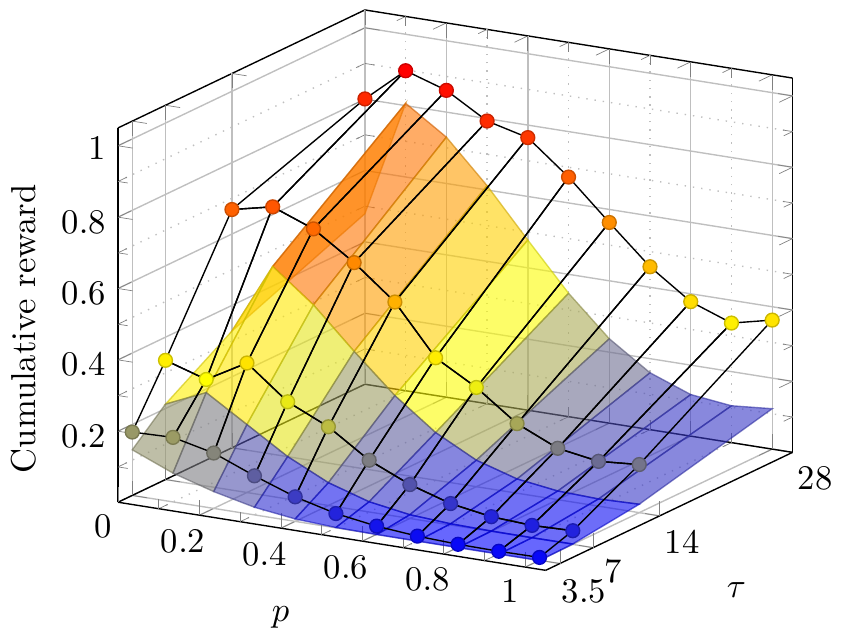}
    \caption{Cumulative reward of a Neural Network $\mathcal{N}$ trained for $p'=0,\ \tau'=14$ and then tested on all the $(p, \tau)$ grid (black dotted mesh). The solid colored surface gives the cumulative reward of the free evolution.}
    \label{fig:same_NN_over_grid}
\end{figure}

\begin{figure}
    \centering
    \includegraphics[width=\linewidth]{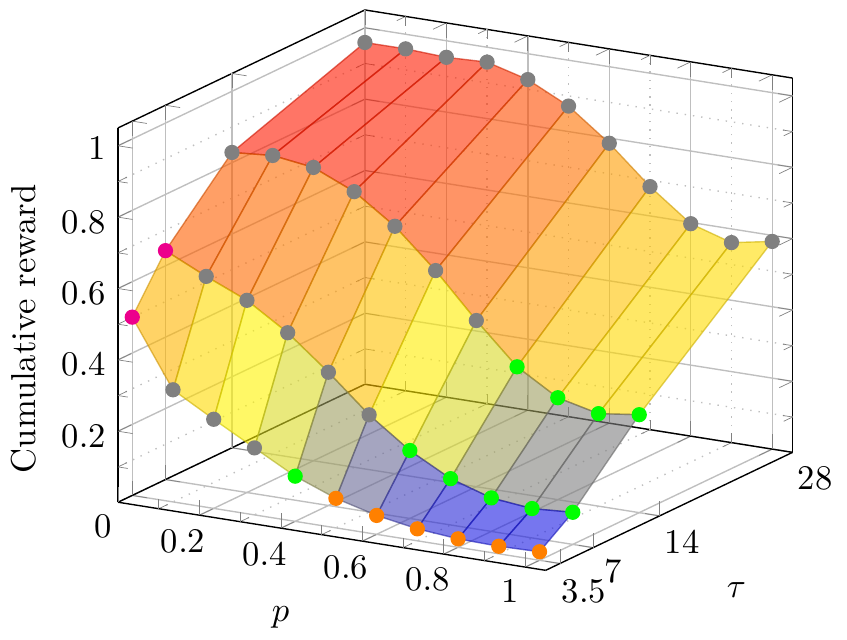}
    \caption{Cumulative reward obtained maximizing for each $(p, \tau)$ the cumulative reward from all the optimal sequences suggested by all the trained Neural Networks. Of all the sequences, four are sufficient to give the maximum cumulative reward, and are identified by the colored markers (gray, magenta, green and orange).
    }
    \label{fig:reward_optimal_sequences}
\end{figure}

After training the learning algorithm, we have verified its generalization properties on unseeing parameter pairs $(p,\ \tau)$. Namely, we have applied the Neural Network $\mathcal{N}$ trained for $(p',\tau')$ on all the $(p,\ \tau)$ grid. An example of the cumulative reward obtained from this comparison is depicted in Fig. \ref{fig:same_NN_over_grid}, where we represent with a colored surface the performance of the free evolution, and in a black mesh surface the cumulative reward of the Neural Network trained for $p'=0,\ \tau'=14$. The figure, to be compared with Fig. \ref{fig:surface}, is qualitatively similar, meaning that a single Neural Network is able to generalize the behaviour of other Networks trained for each $(p,\ \tau)$. Note that the optimal sequences proposed by $\mathcal{N}$ are indeed different depending on $(p,\ \tau)$ (though they may share similar patterns), and that in general the optimal sequences of actions are optimal only locally. We have tested this latter hypothesis running all the optimal sequences proposed by all the trained Neural Networks for all grid points $(p,\ \tau)$. The cumulative reward obtained is plotted in Fig. \ref{fig:reward_optimal_sequences}, where we can observe that a small number of optimal sequences cover all the grid (the same sequence is related to the same color of the marker). Despite not being an exhaustive check for all the possible sequences, this gives evidence that a sequence is optimal only locally.

\subsection*{Robustness}

\begin{figure}[htbp!]
    \centering
    \includegraphics[width=\linewidth]{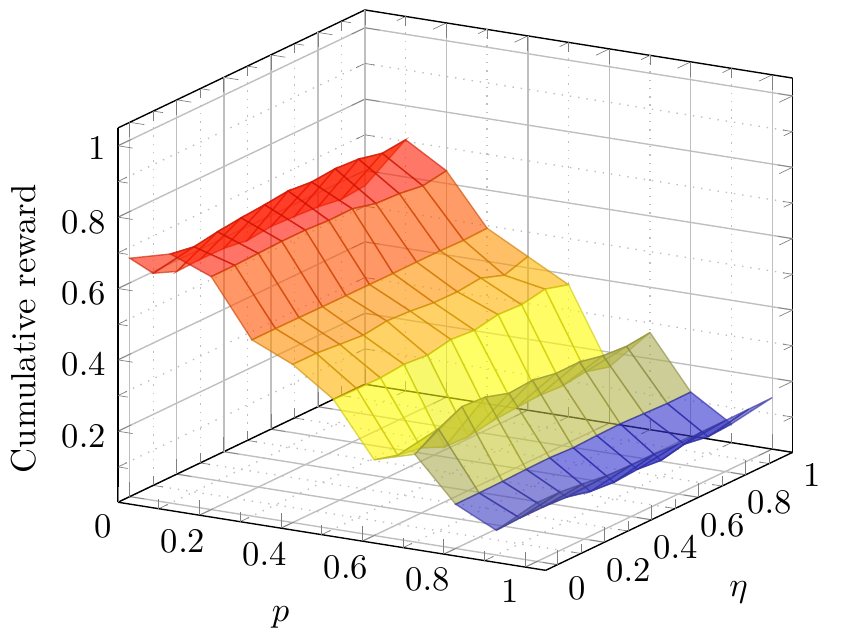}
    \caption{Cumulative reward of an agent trained at $\tau=14$ and deployed in a noisy environment where the noise is controlled by the parameter $\eta$. The reported reward is the average performance of the agent over $100$ different realizations of the noise.}
    \label{fig:noise_time}
\end{figure}
To further test the robustness of our trained agent we checked its performances in a stochastic environment where the time interval between the actions can fluctuate. The agent is thus forced to adapt its strategy to the new environment and, as we can observe, in Fig.~\ref{fig:noise_time} it does this surprisingly well. The agents have been first trained in a noiseless environment with $\tau=14$ and $p \in [0,1]$.
The additional noise in the time is controlled by a parameter $0 \leq \eta \leq 1$ while the total time of the actions is kept fixed. In detail, for a set of $N=8$ actions we sample 8 random numbers in the interval $[-\eta \tau,\ \eta \tau]$ obtaining a noise vector $\bar{\eta}$, which is then averaged to zero in order to keep the total time of the walker constant.
This vector gives the variation to apply to each time instants $t_k$ where the actions are performed.
In Fig.~\ref{fig:noise_time} we plot the average reward obtained by the agent in this noisy environment over $100$ different realizations of the noise. We can see that as we increase the noise parameter $\eta$ our agent keeps the ability to find the correct actions in order to make the reward consistent even in a noisy environment, even though it has not been retrained in the noisy setting.
This analysis on the robustness to noise in time further proves the capability of our approach to generalize well to different environments.

\subsection*{RL actions timing}

\begin{figure}[t!]
    \centering
    \subfloat[\label{fig:transientPre}]{
        \includegraphics[width=\linewidth]{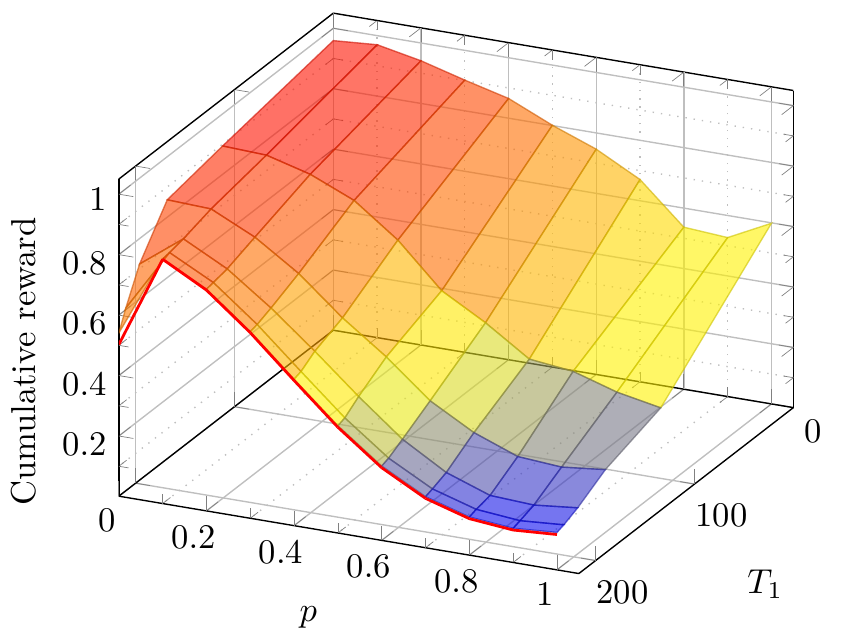}
    }
    \\[2mm]
    \subfloat[\label{fig:transientPost}]{
        \includegraphics[width=\linewidth]{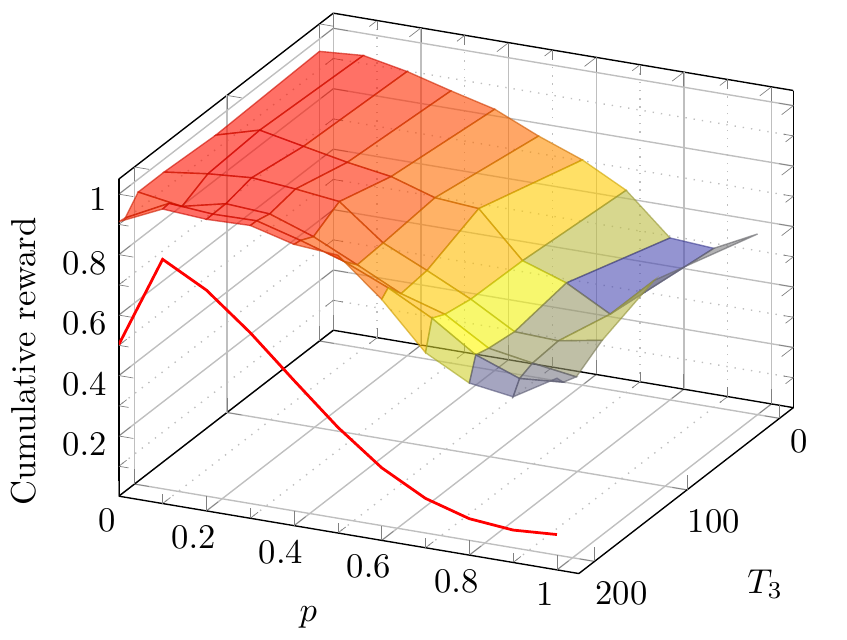}
    }
    \caption{Cumulative reward for a fixed total time evolution $T=T_1+T_2+T_3=8\times 28=224$, with $T_1$ the free evolution interval before the application of the actions, $T_2=8\times \tau$ the interval where the actions are performed, $T_3$ the free evolution interval after the actions. The time unit is given in terms of the inverse of the sink rate set to $1$. The cumulative reward for the no-actions strategy (only free evolution in $T$) is drawn in a red solid line.
    (a)~Cumulative reward as a function of $p$ and $T_1$ with $T_3=0$. Notice that in the limit where $T_1=224$ the actions are packed at the end of the time evolution where they become irrelevant, thus recovering the no-action case.
    (b)~Cumulative reward as a function of $p$ and $T_3$ with $T_1=0$.
    }
    \label{fig:transient}
\end{figure}

Finally, we analyze the scenario when one introduces a transient time before or after the set of equally-time-spaced actions. Namely, we consider a total time evolution of $T=8\times 28=224$, and split it in $T=T_1+T_2+T_3$ where $T_1$ is a transient time with free evolution before applying the actions, $T_2=N\times \tau$ is the time interval applying the actions spaced by $\tau$, and $T_3$ is a transient time of free evolution after the actions. The results in Fig. \ref{fig:transient} show that our training method is applicable also in this more general scenario, and we can also observe the role of the time instant to perform the action. In fact, accumulating the actions at the beginning (Fig. \ref{fig:transient}b) and at the end (Fig. \ref{fig:transient}a) of the dynamics seems to lead to a suboptimal strategy, where the improvements are more difficult to occur. Of course, the extreme case of performing the actions at the very end shows no improvement with respect to the no-action strategy (Fig. \ref{fig:transient}a for large $T_1$). We also find that for low values $p$ (meaning more quantumness of the walker) the time at which we do the actions is clearly less important than for large values of $p$, where a different timing of the actions can result in a drastic reduction of the improvements over the baseline. This results proves the robustness of the quantum regime with respect to the classical one.

\newpage
\bibliographystyle{unsrt}
\bibliography{bibliography}

\begin{thebibliography}{10}

\bibitem{bishop_pattern_2011}
Christopher~M. Bishop.
\newblock {\em Pattern recognition and machine learning}.
\newblock Springer, New York, 1st ed. 2006. corr. 2nd printing 2011 edition
  edition, April 2011.

\bibitem{cover_elements_1991}
T.~M. Cover and Joy~A. Thomas.
\newblock {\em Information theory and statistics}.
\newblock Wiley series in telecommunications. Wiley, New York, 1991.

\bibitem{hastie2009elements}
Trevor Hastie, Robert Tibshirani, and Jerome Friedman.
\newblock {\em The elements of statistical learning: data mining, inference,
  and prediction}.
\newblock Springer Science \& Business Media, 2009.

\bibitem{Sutton2018}
Richard~S Sutton and Andrew~G Barto.
\newblock {\em Reinforcement learning: An introduction}.
\newblock MIT press, 2018.

\bibitem{mnih2015human}
Volodymyr Mnih, Koray Kavukcuoglu, David Silver, Andrei~A Rusu, Joel Veness,
  Marc~G Bellemare, Alex Graves, Martin Riedmiller, Andreas~K Fidjeland, Georg
  Ostrovski, et~al.
\newblock Human-level control through deep reinforcement learning.
\newblock {\em Nature}, 518(7540):529--533, 2015.

\bibitem{Silver2016}
David Silver, Aja Huang, Chris~J. Maddison, Arthur Guez, Laurent Sifre, George
  van~den Driessche, Julian Schrittwieser, Ioannis Antonoglou, Veda
  Panneershelvam, Marc Lanctot, Sander Dieleman, Dominik Grewe, John Nham, Nal
  Kalchbrenner, Ilya Sutskever, Timothy Lillicrap, Madeleine Leach, Koray
  Kavukcuoglu, Thore Graepel, and Demis Hassabis.
\newblock Mastering the game of {Go} with deep neural networks and tree search.
\newblock {\em Nature}, 529(7587):484--489, jan 2016.

\bibitem{kiran2020deep}
B~Ravi Kiran, Ibrahim Sobh, Victor Talpaert, Patrick Mannion, Ahmad A~Al
  Sallab, Senthil Yogamani, and Patrick P{\'e}rez.
\newblock Deep reinforcement learning for autonomous driving: A survey.
\newblock {\em arXiv preprint arXiv:2002.00444}, 2020.

\bibitem{botvinick2020deep}
Matthew Botvinick, Jane~X Wang, Will Dabney, Kevin~J Miller, and Zeb
  Kurth-Nelson.
\newblock Deep reinforcement learning and its neuroscientific implications.
\newblock {\em Neuron}, 2020.

\bibitem{schuld2015introduction}
Maria Schuld, Ilya Sinayskiy, and Francesco Petruccione.
\newblock An introduction to quantum machine learning.
\newblock {\em Contemporary Physics}, 56(2):172--185, 2015.

\bibitem{wittek2014quantum}
Peter Wittek.
\newblock {\em Quantum machine learning: what quantum computing means to data
  mining}.
\newblock Academic Press, 2014.

\bibitem{adcock2015advances}
Jeremy Adcock, Euan Allen, Matthew Day, Stefan Frick, Janna Hinchliff, Mack
  Johnson, Sam Morley-Short, Sam Pallister, Alasdair Price, and Stasja
  Stanisic.
\newblock Advances in quantum machine learning.
\newblock {\em arXiv preprint arXiv:1512.02900}, 2015.

\bibitem{arunachalam2017survey}
Srinivasan Arunachalam and Ronald de~Wolf.
\newblock A survey of quantum learning theory.
\newblock {\em arXiv preprint arXiv:1701.06806}, 2017.

\bibitem{Biamonte:2017db}
Jacob Biamonte, Peter Wittek, Nicola Pancotti, Patrick Rebentrost, Nathan
  Wiebe, and Seth Lloyd.
\newblock Quantum machine learning.
\newblock {\em Nature}, 549:195--202, 2017.

\bibitem{neven2008training}
Hartmut Neven, Vasil~S Denchev, Geordie Rose, and William~G Macready.
\newblock Training a binary classifier with the quantum adiabatic algorithm.
\newblock {\em arXiv preprint arXiv:0811.0416}, 2008.

\bibitem{mott2017solving}
Alex Mott, Joshua Job, Jean-Roch Vlimant, Daniel Lidar, and Maria Spiropulu.
\newblock Solving a higgs optimization problem with quantum annealing for
  machine learning.
\newblock {\em Nature}, 550(7676):375, 2017.

\bibitem{lloyd20}
Seth Lloyd, Maria Schuld, Aroosa Ijaz, Josh Izaac, and Nathan Killoran.
\newblock Quantum embeddings for machine learning.
\newblock {\em arXiv preprint arXiv:2001.03622}, 2020.

\bibitem{martina2021machine}
Stefano Martina, Stefano Gherardini, and Filippo Caruso.
\newblock Machine learning approach for quantum non-markovian noise
  classification.
\newblock {\em arXiv preprint arXiv:2101.03221}, 2021.

\bibitem{otterbach2017unsupervised}
JS~Otterbach, R~Manenti, N~Alidoust, A~Bestwick, M~Block, B~Bloom, S~Caldwell,
  N~Didier, E~Schuyler Fried, S~Hong, et~al.
\newblock Unsupervised machine learning on a hybrid quantum computer.
\newblock {\em arXiv preprint arXiv:1712.05771}, 2017.

\bibitem{vinci2020path}
Walter Vinci, Lorenzo Buffoni, Hossein Sadeghi, Amir Khoshaman, Evgeny
  Andriyash, and Mohammad Amin.
\newblock A path towards quantum advantage in training deep generative models
  with quantum annealers.
\newblock {\em Machine Learning: Science and Technology}, 2020.

\bibitem{hu2019quantum}
Ling Hu, Shu-Hao Wu, Weizhou Cai, Yuwei Ma, Xianghao Mu, Yuan Xu, Haiyan Wang,
  Yipu Song, Dong-Ling Deng, Chang-Ling Zou, et~al.
\newblock Quantum generative adversarial learning in a superconducting quantum
  circuit.
\newblock {\em Science advances}, 5(1):eaav2761, 2019.

\bibitem{Zonghai2005}
Chen~Z. Dong~D., Chen~C.
\newblock Quantum reinforcement learning.
\newblock {\em Notes in Computer Science}, 3611:686--689, 2005.

\bibitem{Paparo2014}
Giuseppe~Davide Paparo, Vedran Dunjko, Adi Makmal, Miguel~Angel Martin-Delgado,
  and Hans~J. Briegel.
\newblock Quantum speedup for active learning agents.
\newblock {\em Phys. Rev. X}, 4:031002, Jul 2014.

\bibitem{Dunjko2016}
Vedran Dunjko, Jacob~M. Taylor, and Hans~J. Briegel.
\newblock Quantum-enhanced machine learning.
\newblock {\em Physical Review Letters}, 117(13):130501, September 2016.

\bibitem{Saggio2021}
Valeria Saggio, Beate~E Asenbeck, Arne Hamann, Teodor Str{\"o}mberg, Peter
  Schiansky, Vedran Dunjko, Nicolai Friis, Nicholas~C Harris, Michael Hochberg,
  Dirk Englund, et~al.
\newblock Experimental quantum speed-up in reinforcement learning agents.
\newblock {\em Nature}, 591(7849):229--233, 2021.

\bibitem{Breuer2002}
Heinz-Peter Breuer and Francesco Petruccione.
\newblock {\em The theory of open quantum systems}.
\newblock Oxford University Press, 2002.

\bibitem{CarusoRMP2014}
Filippo Caruso, Vittorio Giovannetti, Cosmo Lupo, and Stefano Mancini.
\newblock Quantum channels and memory effects.
\newblock {\em Reviews of Modern Physics}, 86(4):1203, 2014.

\bibitem{caruso2009highly}
Filippo Caruso, Alex~W Chin, Animesh Datta, Susana~F Huelga, and Martin~B
  Plenio.
\newblock Highly efficient energy excitation transfer in light-harvesting
  complexes: The fundamental role of noise-assisted transport.
\newblock {\em The Journal of Chemical Physics}, 131(10):09B612, 2009.

\bibitem{Caruso2016}
Filippo Caruso, Andrea Crespi, Anna~Gabriella Ciriolo, Fabio Sciarrino, and
  Roberto Osellame.
\newblock Fast escape of a quantum walker from an integrated photonic maze.
\newblock {\em Nature Communications}, 7:11682, June 2016.

\bibitem{Ghavamzadeh2015}
Mohammed Ghavamzadeh, Shie Mannor, Joelle Pineau, and Aviv Tamar.
\newblock Bayesian reinforcement learning: A survey.
\newblock {\em Foundations and Trends in Machine Learning}, 8(5-6):359--483,
  2015.

\bibitem{Whitfield2010}
James~D. Whitfield, C\'esar~A. Rodr\'{\i}guez-Rosario, and Al\'an Aspuru-Guzik.
\newblock Quantum stochastic walks: A generalization of classical random walks
  and quantum walks.
\newblock {\em Phys. Rev. A}, 81:022323, Feb 2010.

\bibitem{Caruso2014}
Filippo Caruso.
\newblock Universally optimal noisy quantum walks on complex networks.
\newblock {\em New Journal of Physics}, 16(5):055015, 2014.

\bibitem{DallaPozza2020}
Nicola Dalla~Pozza and Filippo Caruso.
\newblock Quantum state discrimination on reconfigurable noise-robust quantum
  networks.
\newblock {\em Phys. Rev. Research}, 2:043011, Oct 2020.

\bibitem{lindblad1976generators}
Goran Lindblad.
\newblock On the generators of quantum dynamical semigroups.
\newblock {\em Communications in Mathematical Physics}, 48(2):119--130, 1976.

\bibitem{johansson2012qutip}
J~Robert Johansson, Paul~D Nation, and Franco Nori.
\newblock Qutip: An open-source python framework for the dynamics of open
  quantum systems.
\newblock {\em Computer Physics Communications}, 183(8):1760--1772, 2012.

\bibitem{brockman2016openai}
Greg Brockman, Vicki Cheung, Ludwig Pettersson, Jonas Schneider, John Schulman,
  Jie Tang, and Wojciech Zaremba.
\newblock Openai gym.
\newblock {\em arXiv preprint arXiv:1606.01540}, 2016.

\bibitem{Stooke2019}
Adam Stooke and Pieter Abbeel.
\newblock rlpyt: A research code base for deep reinforcement learning in
  pytorch.
\newblock {\em arXiv preprint arXiv:1909.01500}, 2019.

\bibitem{PyTorch}
Adam Paszke, Sam Gross, Francisco Massa, Adam Lerer, James Bradbury, Gregory
  Chanan, Trevor Killeen, Zeming Lin, Natalia Gimelshein, Luca Antiga, Alban
  Desmaison, Andreas Kopf, Edward Yang, Zachary DeVito, Martin Raison, Alykhan
  Tejani, Sasank Chilamkurthy, Benoit Steiner, Lu~Fang, Junjie Bai, and Soumith
  Chintala.
\newblock Pytorch: An imperative style, high-performance deep learning library.
\newblock In H.~Wallach, H.~Larochelle, A.~Beygelzimer, F.~Alch\'{e}-Buc,
  E.~Fox, and R.~Garnett, editors, {\em Advances in Neural Information
  Processing Systems 32}, pages 8024--8035. Curran Associates, Inc., 2019.

\bibitem{Hyperopt}
James Bergstra, Daniel Yamins, and David Cox.
\newblock Making a science of model search: Hyperparameter optimization in
  hundreds of dimensions for vision architectures.
\newblock In Sanjoy Dasgupta and David McAllester, editors, {\em Proceedings of
  the 30th International Conference on Machine Learning}, volume~28 of {\em
  Proceedings of Machine Learning Research}, pages 115--123, Atlanta, Georgia,
  USA, 17--19 Jun 2013. PMLR.

\bibitem{liaw2018tune}
Richard Liaw, Eric Liang, Robert Nishihara, Philipp Moritz, Joseph~E Gonzalez,
  and Ion Stoica.
\newblock Tune: A research platform for distributed model selection and
  training.
\newblock {\em arXiv preprint arXiv:1807.05118}, 2018.

\end{thebibliography}

\end{document}